\documentclass[preprint,prd,showpacs,showkeys]{revtex4}
\usepackage{graphicx}
\usepackage{color}
\usepackage{amsmath}
\usepackage{amssymb, graphics}
\usepackage{latexsym}
\usepackage{bm}
\usepackage{wrapfig}
\usepackage{fancybox}
\def\be{\begin{equation}}
\def\ee{\end{equation}}
\def\bea{\begin{eqnarray}}
\def\eea{\end{eqnarray}}
\begin{document}

\title{On the ensemble dependence in black hole geometrothermodynamics}

\author{Hernando Quevedo$^{1,2}$,  Alberto S\'anchez$^3$ and Safia Taj$^{4}$}
\email{quevedo@nucleares.unam.mx, asanchez@nucleares.unam.mx, safiataaj@gmail.com}
\affiliation{
$^1$Instituto de Ciencias Nucleares, Universidad Nacional Aut\'onoma de M\'exico, AP 70543,
M\'exico, DF 04510, Mexico\\
$^2$ Instituto de Cosmologia, Relatividade e Astrofisica and ICRANet - CBPF,
Rua Dr. Xavier Sigaud, 150, CEP 22290-180, Rio de Janeiro, Brazil\\
$^3$Departamento de Posgrado, CIIDET, AP 752, Quer\'etaro, QRO 76000, Mexico\\
$^4$Department of Basic Sciences and Humanities, College of
Electrical and Mechanical Engineering, Peshawar Road, Rawalpindi,
Pakistan }

\date{\today}


\begin{abstract}

We investigate the dependence of thermodynamic properties of black
holes on the choice of statistical ensemble for a particular class
of Einstein-Maxwell-Gauss-Bonnet black holes with cosmological
constant. We use partial Legendre transformations in the
thermodynamic limit in order to compare the results in different
ensembles, and show that the phase transition structure  depend on
the choice of thermodynamic potential. This result implies that
thermodynamic metrics which are partially Legendre invariant cannot
be employed to describe black hole thermodynamics, and partly explains  
why a particular thermodynamic metric has been used so far in the framework of black hole
geometrothermodynamics.


\end{abstract}

\maketitle

\section{Introduction}
\label{sec:int}

In the Euclidean path-integral approach to black hole
thermodynamics, originally developed by Hawking et al.
\cite{haw76,hh76,gh77,haw79}, the thermodynamic partition function
is computed from the path integral in the saddle-point
approximation, obtaining as a result the laws of black hole
thermodynamics. Originally, only the microcanonical ensemble was
investigated \cite{haw76,haw79} because the canonical ensemble led
to difficulties related to the stability of the black hole under
consideration. Later on, the canonical ensemble was investigated by
York et al. \cite{york86,york89,wy88,whi90,by94,bbwy90} by using
appropriate boundary conditions. It turns out that the Euclidean
path-integral approach is not well defined until the boundary
conditions of the system are properly defined. The various ensembles
available to the system are defined by identifying the corresponding
boundary conditions. York's approach was used to analyze other
ensembles \cite{brown90} and to study particular systems like the
charged black hole in the grand canonical ensemble \cite{bbwy90},
black holes in asymptotically anti-de Sitter spacetimes
\cite{bcm94,louwin96}, and black holes in two and three dimensions
\cite{lem96,bcm94}.

Under certain circumstances, the results obtained by using the
path-integral approach turned out to depend on the boundary
conditions \cite{bcm94,hawpag83,page92}. The stability of black
holes also turned out to depend on the choice of boundary conditions
and, consequently, on the ensemble \cite{comer92}. In fact, it was
found that in one ensemble the black hole can never be stable,
independently of the values of the black hole parameters, and in
another ensemble the black hole is almost always stable. This result
shows that the stability properties of a black hole are drastically
influenced by the boundary conditions that determine ensemble.
Moreover, it follows that this behavior must be also present in the
thermodynamic limit in which different ensembles  correspond to
different thermodynamic potentials, related by Legendre
transformations. On the other hand, an important characteristic of a
black hole is its phase transition structure which is closely
related to the stability properties of the system. It then follows
that the phase transition structure can, in principle,  also be
ensemble dependent. In the thermodynamic limit, this would imply a
dependence on the thermodynamic potential.

The formalism of geometrothermodynamics (GTD) has been proposed in
\cite{quev07} as a Legendre invariant approach to describe
thermodynamics in terms of geometric concepts. One of the conjectures of
GTD is that curvature singularities of the equilibrium space are
related to phase transitions of the system. One can then wonder
whether GTD is able to handle the dependence of phase transition
structure on the statistical ensemble. This is the main goal of the
present work. For the sake of concreteness, we will consider here a
particular black hole configuration in the
Einstein-Maxwell-Gauss-Bonnet (EMGB) theory. We will see that in
fact the phase transition structure, as dictated by the behavior of
the specific heats, drastically depends on the thermodynamic
potentials which are related by partial Legendre transformations.
Consequently, the metrics that in GTD are invariant under partial
Legendre transformations cannot be used to describe in an invariant
manner the properties of such black hole configurations.

This paper is organized as follows. In Sec. \ref{sec:gtd}, we
present a brief introduction into the formalism of GTD. In Sec.
\ref{sec:emgbl}, we review the main aspects of a particular
spherically symmetric black hole in the EMGB theory with
cosmological constant. In Sec. \ref{sec:gtdana}, we apply the
formalism of GTD with a particular thermodynamic metric to reproduce
the thermodynamic properties of the black hole configuration.
Finally, in Sec. \ref{sec:con}, we discuss our results.


\section{Geometrothermodynamics}
\label{sec:gtd}

In equilibrium thermodynamics, to describe a system with $n$
thermodynamic degrees of freedom, one needs a thermodynamic
potential $\Phi$, a set of $n$ extensive variables $\{E^a\}$
($a=1,...,n$), and the corresponding dual intensive variables
$\{I^a\}$. Classical thermodynamics is invariant with respect to a
change of thermodynamic potential $\Phi\rightarrow \tilde\Phi$ which
is defined by meas of the Legendre transformations \cite{arnold} \be
\{Z^A\}\longrightarrow \{\widetilde{Z}^A\}=\{\tilde \Phi, \tilde E
^a, \tilde I ^ a\}\ , \ee \be
 \Phi = \tilde \Phi - \delta_{kl} \tilde E ^k \tilde I ^l \ ,\quad
 E^i = - \tilde I ^ {i}, \ \
E^j = \tilde E ^j,\quad
 I^{i} = \tilde E ^ i , \ \
 I^j = \tilde I ^j \ ,
 \label{leg}
\ee where $i\cup j$ is any disjoint decomposition of the set of
indices $\{1,...,n\}$, and $k,l= 1,...,i$. In particular, for
$i=\emptyset$ we obtain the identity transformation. Moreover, for
$i=\{1,...,n\}$, Eq.(\ref{leg}) defines a total Legendre
transformation, i.e., \be
 \Phi = \tilde \Phi - \delta_{ab} \tilde E ^a \tilde I ^b \ ,\quad
 E^a = - \tilde I ^ {a}, \ \
 I^{a} = \tilde E ^ a \ .
 \label{legtotal}
\ee Notice that in order to apply a Legendre transformation to a
tensorial object in a particular coordinate system,  it is necessary
to use  the corresponding matrix representation which can be
computed by considering all the coordinates as independent, that is,
as coordinates of a $(2n+1)-$dimensional space \cite{quev07}. It
then turns out that to investigate  the mathematical structure of
thermodynamics in general it is necessary to use contact geometry
which is constructed as follows. Consider a $(2n+1)-$dimensional
differential manifold ${\cal T}$ together with its tangent manifold
$T({\cal T})$. Let ${\cal V}\subset T({\cal T})$ be an arbitrary
field of hyperplanes on ${\cal T}$. It can be shown that there
exists a non-vanishing differential 1-form $\Theta$ on the cotangent
manifold $T^*({\cal T})$ such that ${\cal V} = \ker \Theta$. If the
Frobenius integrability condition $\Theta\wedge d \Theta =0$ is
satisfied, the hyperplane field ${\cal V}$ is said to be completely
integrable. On the contrary, if   $\Theta \wedge d \Theta \neq 0$,
then ${\cal V}$ is non-integrable. In the limiting case $\Theta
\wedge (d \Theta)^n \neq 0$, the hyperplane field ${\cal V}$ becomes
maximally non-integrable and  is said to define a contact structure
on ${\cal T}$. The pair $({\cal T}, \Theta)$ determines a contact
manifold \cite{handbook}. Consider $G$ as a non-degenerate metric on
${\cal T}$. The set $({\cal T}, \Theta, G)$ defines a Riemannian
contact manifold.

Let us choose the coordinates of ${\cal T}$ as $Z^A =\{\Phi, E^a, I ^a\}$
 with $A=0,1,...,2n$. According to Darboux theorem,
the contact 1--form can be written as
\be
 \Theta= d\Phi - \delta_{ab} I^a d E^b \ ,\quad \delta_{ab}={\rm diag} (1,1,...,1)\ ,
\label{gibbs} \ee where we assume the convention of summation over
repeated indices. Under a Legendre transformation, the contact
1-form transforms as $\Theta\rightarrow \tilde\Theta =  d\tilde \Phi
- \delta_{ab} \tilde I^a d \tilde E^b$. Consequently, the contact
manifold $({\cal T}, \Theta)$ is a Legendre invariant structure. If
we now impose Legendre invariance on the metric $G$, the Riemannian
contact manifold $({\cal T}, \Theta,G)$ is Legendre invariant. Any
Riemannian contact manifold $({\cal T}, \Theta,G)$ which satisfies
the condition of Legendre invariance is called a thermodynamic {\it
phase manifold} and constitutes the starting point for a description
of thermodynamic systems in terms of geometric concepts.

Notice that to construct a concrete phase manifold we only need to specify
the metric $G$. It turns out that Legendre invariance does not fix
completely the metric $G$. All the metrics we have found so far can
be classified as invariant under total Legendre transformations
\begin{equation}
G^{I/II} = \left(d\Phi - I_a dE^a\right)^2  +\Lambda
\left(\xi_{ab}E^{a}I^{b}\right)\left(\chi_{cd}dE^{c}dI^{d}\right) \  ,
\label{gup1}
\end{equation}
or invariant also under partial Legendre transformations \be
\label{ginv2} G^{III}=\left(d\Phi - I_a dE^a\right)^2 + \Lambda\,
(E_a I_a)^{2k+1} d E^a d I^a \ ,\quad E_a =\delta_{ab}E^b\ ,\quad
I_a =\delta_{ab} I ^b\ . \ee Here $\Lambda$ is an arbitrary Legendre
invariant real function of $Z^A$, and $\xi_{ab}$ and $\chi_{ab}$ are
diagonal constant matrices that can be expressed in terms of the
Euclidean and pseudo-Euclidean metrics $\delta_{ab}={\rm
diag}(1,...,1)$ and $\eta_{ab} = {\rm diag}(-1, 1, ..., 1)$,
respectively. The choice $\xi_{ab}=\delta_{ab}$,
$\chi_{ab}=\delta_{ab}$ leads to the metric $G^I$ which has been
used to describe systems with first order phase transitions
\cite{quev07,qstv10a}. The alternative choice $\xi_{ab}=\delta_{ab}$
$\chi_{ab}=\eta_{ab}$, which is denoted by $G^{II}$, turned out to
describe correctly second order phase transitions especially in
black hole thermodynamics \cite{qstv10a,aqs08,vqs09}. Moreover, in
the metric $G^{II}$ the additional choice $\xi_{ab}=
\frac{1}{2}\left(\delta_{ab}-\eta_{ab}\right)$ can be used to handle
also the thermodynamic limit $T\rightarrow 0$.

In classical thermodynamics, a particular thermodynamic system is
completely determined by its fundamental equation $\Phi=\Phi(E^a)$
which determines an $n-$dimensional surface in ${\cal T}$. In GTD,
we use the same idea to construct the manifold which should describe
a particular thermodynamic system. However, some technical details
must be considered in order to preserve the Legendre invariance of
${\cal T}$. To this end, consider a (smooth) embedding  map
$\varphi: {\cal E}\rightarrow {\cal T}$, where ${\cal E}$ is an
$n-$dimensional submanifold  of the phase manifold $({\cal
T},\Theta,G)$. If we consider the set $\{E^a\}$ as the coordinates
of ${\cal E}$, then the embedding map reads $ \varphi :  \{E^a\}
\longmapsto \{Z^A(E^a)\}=\{\Phi(E^a), E^a, I^a(E^a)\}$. In this
manner, the fundamental equation $\Phi=\Phi(E^a)$ appears in a
natural way as the result of introducing a well-defined embedding
map $\varphi$. Moreover, the metric $G$ induces in ${\cal E}$ the
canonical metric $g$ by means of $g=\varphi^*(G)$, where $\varphi^*$
is the pullback of $\varphi$. The pair $({\cal E},g)$ is called
equilibrium manifold if the  map $\varphi: {\cal E}\rightarrow {\cal
T}$ satisfies the condition \be \varphi^*(\Theta)=\varphi^*(d\Phi
-\delta_{ab}\, I^a\, dE^b) = 0\ , \ee which implies that
\begin{equation}
 d\Phi
=I_{a}dE^{a}\ , \quad \frac{\partial \Phi}{\partial E^{a}}=I_{a}\ .
\label{firstlaw}
\end{equation}
The first of these equations corresponds to the first law of
thermodynamics whereas the second one is usually known as the
condition for thermodynamic equilibrium \cite{callen}.

In the case of the metric $G^{II}$ discussed above with
$\xi_{ab}=\frac{1}{2}\left(\delta_{ab}-\eta_{ab}\right)$ and
$\chi_{ab}=\eta_{ab}$, the corresponding induced metric of ${\cal
E}$ can be written as \be g^{II}=\frac{1}{2} \Lambda   \left(
E^d\frac{\partial\Phi}{\partial E^d}  - \eta_d^{c}E^d\frac{\partial
\Phi}{\partial E^c}\right) \left(\eta_a^{b} \frac{\partial^2
\Phi}{\partial E^b \partial E^c}
 dE^a d E^c\right)\ ,
\label{gdown} \ee where $\eta^{b}_{a} ={\rm diag}(-1,1,...1)$. The
geometric properties of the equilibrium manifold ${\cal E}$
described by the metric $g^{II}$ should be related to the
thermodynamic properties of the system described by the fundamental
equation $\Phi(E^a)$. In GTD, it is conjectured that ${\cal E}$ is
curved for systems with thermodynamic interaction and that curvature
singularities in ${\cal E}$ correspond to phase transitions of the
corresponding thermodynamic system.

Notice that the arbitrary function $\Lambda$ represents a Legendre
invariant conformal factor. In practice, it does not affect the main
behavior of the curvature of $g^{II}$ which is the main geometric
quantity in GTD. Nevertheless, this function plays an important role
if we demand  that our results be invariant not only under Legendre
transformations, but also with respect to changes of representation.
In fact, one can show \cite{bnlq13} that a particular choice of
$\Lambda$ makes the metric $G^{I}$ invariant under changes of
representation in general. In the case of the metric $G^{II}$, an
additional conformal factor is needed \cite{bmmq13}.

\section{A spherically symmetric charged black hole}
\label{sec:emgbl}

In this section, we study a particular black hole solution which is
appropriate to illustrate the ensemble dependence of black hole
thermodynamics. The Einstein-Gauss-Bonnet (EGB) theory is the most
general theory in five dimensions that leads to second order
differential equations, although the corresponding Lagrangian
density contains quadratic powers of the curvature. The most general
action of the EGB theory is obtained by adding the Gauss-Bonnet (GB)
invariant and the matter Lagrangian $L_{matter}$ to the
Einstein-Hilbert action
\begin{equation}
I= \kappa\int d^{5}x \sqrt{g}[R +\alpha(R^{2}-4R^{\mu\nu}R_{\mu%
\nu}+R^{\alpha\beta\gamma\delta}R_{\alpha\beta\gamma\delta}) + L_{matter}],
\label{egbaction}
\end{equation}
where $\kappa$ is related to the Newton constant, and $\alpha$ is
the GB coupling constant.
In the case of the Einstein--Maxwell-Gauss-Bonnet (EMGB) theory with
cosmological constant, the matter component of the action
(\ref{egbaction}) is given by
\be
 L_{matter}= F_{\alpha\beta}F^{\alpha\beta} - 2 \Lambda
 \ , \quad F_{\alpha\beta}=A_{\beta,\alpha}-A_ {\alpha,\beta}\ ,
\ee
where $\Lambda$ is the cosmological constant, and $F_{\alpha\beta}$ represents
the electromagnetic Faraday tensor.

The low energy limit of certain string theories leads to the
EMGB theory with cosmological constant; therefore, it is
important to study the physical properties of exact solutions like
black hole solutions. A particular solution was obtained in
\cite{A18} (see also \cite{odin1,A21,A22}) by using the following 5D
static spherically symmetric line element
\begin{equation}
ds^{2}= -f(r)dt^{2}+\frac{dr^{2}}{f(r)}+r^{2}[d\theta_{1}^{2}+
\sin^{2}\theta_{1}(d\theta_{2}^{2}+\sin^{2}\theta_{2}
d\theta_{3}^{2}) ] \ .
\label{lelemgb}
\end{equation}%
The coordinate $r$ has the dimension of length while the
angular coordinates $(\theta_{1},\theta_{2})\in[0,\pi]$ and
$\theta_{3} \in[0,2\pi]$. A 5D spherically symmetric solution in EMGB gravity
with $\Lambda$ was obtained by Wiltshire \cite{wilt}, using the
metric ansatz (\ref{lelemgb}) and the metric function
\begin{equation}
f(r)=1+\frac{r^{2}}{4\alpha }-\frac{r^{2}}{4\alpha
}\sqrt{1+\frac{8\alpha M }{ r^{4}}-\frac{8\alpha Q^{2}}{3r^{6}}+\frac{4\alpha
\Lambda }{3}}\ .
\label{fungl}
\end{equation}%
The two parameters $M >0$ and $Q$ are identified as the mass and
electric charge of the system. It is easy to see that the conditions
for the solution (\ref{fungl}) to describe a black hole spacetime
are $f(r_H)=0$ and 
\be 1+\frac{8\alpha M }{ r_H^{4}}-\frac{8\alpha
Q^{2}}{3r_H^{6}}+\frac{4\alpha \Lambda }{3}
> 0\ ,
\ee 
where $r_H$ is the radius of the outermost horizon. 
In this work, we will limit ourselves to the investigation of
positive $\alpha$ and negative definite $\Lambda$ in order for the
mass of the black hole to be always positive.

\section{Geometrothermodynamic analysis}
\label{sec:gtdana}

To find the fundamental equation of the black hole under consideration, we first note
that the surface area of the event horizon is
\begin{equation}
A=r_{H
}^{3}\int_{\theta=0}^{\pi}\int_{\phi=0}^{\pi}\int_{\psi=0}^{2\pi}
{\sin^{2}\theta} \sin\phi d\theta d\phi d\psi=2\pi^{2}r_{H }^{3} \ .
\end{equation}
Moreover, the Bekenstein-Hawking entropy  is
$S=\frac{k_{B}A}{4G\hbar}=\frac {k_{B}\pi ^{2}}{2G\hbar}r_{H }^{3}$,
which becomes $S=r_{H }^{3}$ by choosing the constants appropriately \cite{A19}.
Since the black hole condition $f(r_H)=0$ implies that
\begin{equation}
\frac{\Lambda}{3}r_{H}^{6}-2r^{4}_{H}+2\left(M-2\alpha\right)r_{H}^{2}-
\frac{2}{3} Q^{2}=0\ ,
\end{equation}%
the corresponding thermodynamic fundamental equation in the mass representation becomes
\begin{equation}
M=2 \alpha+ S^{2/3} + \frac{Q^2}{3 S^{2/3}} -\frac{\Lambda}{6} S^{4/3} \ .
\label{feqlam}
\end{equation}%
Notice that to guarantee the positiveness of the mass,  we must
demand  that $\alpha >0$ and $\Lambda < 0$.

From the energy conservation law for black holes, $dM = TdS + \phi
dQ$, we can derive the expressions for the temperature and electric
potential of the black hole on the event horizon as
\begin{equation}
T=\frac{2}{9}\,{\frac {3\,{S}^{4/3}-\Lambda\,{S}^{2}-{Q}^{2}}{{S}^{5/3}}}\ ,\quad
\phi=\frac{2}{3} \frac{ Q}{S^{\frac{2}{3}}}.
\label{temgbl}
\end{equation}%

According to Davies \cite{A20}, a black hole undergoes a second
order phase transition at those points where the specific heat $C$
diverges. Strictly speaking, this means that we must introduce the
concept of ``heat", say $Q_{heat}$, for a black hole. Of course,
this is a problem that can be handled correctly only within the
context of a physically meaningful statistical model which is
probably the most important unsolved problem in black hole
thermodynamics. The simplest available solution of this problem is
to use the analogy with classical thermodynamics. Indeed, the first
law of thermodynamics  $dM = TdS + \phi dQ$ allows us to define the
``heat" through the relationship $dQ_{heat}\equiv TdS$ so that
$dM=dQ_{heat} + \phi dQ$. Then, following the standard approach of
ordinary thermodynamics,  we introduce the specific heat as \be C_Q
\equiv \left(\frac{\partial Q_{heat}}{\partial T}\right)_Q =
\left(\frac{\partial M}{\partial T}\right)_Q \ee which in this case
is given by
\begin{equation}
C_{Q}= 3S\left(\frac{3S^{\frac{4}{3}}-\Lambda
S^{2}-Q^{2}}{5Q^{2}-3S^{\frac{4}{3}}-\Lambda S^{2}}\right) \ .
\label{CQeymgb}
\end{equation}%

The temperature (\ref{temgbl})  is positive only in the range
$3\,{S}^{4/3}-\Lambda\,{S}^{2}> {Q}^{2}$ and, therefore, the
specific heat can take either positive
$(5Q^{2}-3S^{\frac{4}{3}}-\Lambda S^{2}>0)$ or negative
$(5Q^{2}-3S^{\frac{4}{3}}-\Lambda S^{2}<0)$ values, indicating the
possibility of stable and unstable states. In the limiting case
$5Q^{2}-3S^{\frac{4}{3}}-\Lambda S^{2}=0$, the black undergoes a
second order phase transition during which it changes its state of
thermodynamic stability.

In classical thermodynamics, once the internal energy of the system
is well defined, the analysis at the level of thermodynamic
variables can be  associated with a particular statistical ensemble.
In the case of black holes, however, there is no unique definition
of internal energy. For instance, the potential $M(S,Q)$ has been
associated with the microcanonical ensemble \cite{wei09} and with
the canonical ensemble \cite{myung08}. In the path-integral method,
in which the ensemble is fixed through the boundary conditions, one
could also associate this potential with the grand canonical
ensemble \cite{peclem99}. Using this last option, we can say that
the phase transition structure we have found above for the EMGB
black hole can be associated with the grand canonical ensemble. Now
we can consider the ensemble dependence already found in black hole
thermodynamics. Indeed, since the stability properties of a black
hole can drastically change from one ensemble to another, in the
thermodynamic limit, in which the change of ensemble corresponds to
a change of thermodynamic potential, the same effect can occur.

In the case of the EMGB black hole we are considering here,
performing Legendre transformations on $M(S,Q)$ one can derive  the
potentials \be H\equiv M-\phi Q\ , \quad F\equiv M - TS \ ,\quad
G\equiv M - TS - \phi Q \ , \ee that are known as the enthalpy, the
Helmholtz free energy, and the Gibbs free energy, respectively. The
enthalpy \be H = M-\phi Q=2 \alpha+ S^{2/3} - \frac{3}{4} \phi^2
S^{2/3} -\frac{\Lambda}{6} S^{4/3} \ . \label{feqlm1} \ee satisfies
the first law of thermodynamics, $dH = TdS - Q d\phi$, and can be
associated with the canonical ensemble. Then, if we assume again
that the ``heat" of the black hole is defined by $dQ_{heat}=TdS$,
the specific heat  for fixed $\phi$ is given by \be C_\phi \equiv
\left(\frac{\partial Q_{heat}}{\partial T}\right)_\phi =
\left(\frac{\partial H}{\partial T}\right)_\phi =- 3S \frac{
12S^{1/3} - 9 \phi^2 S^{1/3} - 4\Lambda S}{12S^{1/3} - 9 \phi^2
S^{1/3} + 4\Lambda S} \ . \label{cphilambda} \ee Taking into account
the condition $12S^{1/3} - 9 \phi^2 S^{1/3} - 4\Lambda S >0$, which
follows from the condition $T>0$, we find that stable states
$(C_\phi>0)$ are allowed for entropies in the range $ S^{2/3} >
\frac{3}{4|\Lambda|} (4 - 3 \phi^2)$. Furthermore, for $\phi^2>4/3$
and $S>0$ the last condition is always satisfied, indicating that
stable states always exist in this case. Moreover, the specific heat
(\ref{cphilambda}) shows that the roots of the equation $12S^{1/3} -
9 \phi^2 S^{1/3} + 4\Lambda S=0$ determine the locations where
second order phase transitions occur. Notice that the singularities
of $C_\phi$ are different from those of $C_Q$; consequently, the
corresponding phase transition structures do not coincide. In the
limiting case $\Lambda \rightarrow 0$, the difference is more
dramatic since the specific heats \be C_{Q}=
3S\left(\frac{3S^{\frac{4}{3}}-
Q^{2}}{5Q^{2}-3S^{\frac{4}{3}}}\right) \ , \quad C_\phi = - 3S \ ,
\ee indicate the existence of phase transitions in the potential
$M(S,Q)$ with no transitions at all in the potential $H(S,\phi)$.
This shows that the thermodynamic properties of this black hole are
not invariant with respect to the partial Legendre transformation
that relates the potentials $M(S,Q)$ and $H(S,\phi)$.

Now we turn to the description of the above results within the
framework of GTD. As mentioned in Sec. \ref{sec:gtd}, we must
choose a metric to derive the geometric properties of the
equilibrium manifold. In general, we have three different options,
namely, $G^I$, $G^{II}$ and $G^{III}$. Since $G^{III}$ is invariant
with respect to partial Legendre transformations, the above result
shows that $G^{III}$ cannot be used to describe the thermodynamics of the
EMGB black hole. It follows that the only invariance that can be
imposed in the case of black holes is with respect to total Legendre
transformations for which we can use the metrics $G^I$ and $G^{II}$.
In our experience, we have seen that $G^{II}$ correctly describes
systems with second order phase transitions and now we want to see
whether it can also correctly handle the dependence on the ensemble.

In the grand canonical ensemble, the fundamental equation $M=M(S,Q)$
is given  in Eq.(\ref{feqlam}). Then, the coordinates of the
equilibrium manifold ${\cal E}$ are $E^a=(S,Q)$ and the
thermodynamic potential is $\Phi=M$. The thermodynamic metric
(\ref{gdown}) can then be written as \be g= \frac{4}{27
S^{4/3}}(3S^{4/3} - \Lambda S^2 - Q^2)\left[ \frac{1}{9S^2}(
3S^{4/3} + \Lambda S^2 - 5 Q^2) dS^2 + dQ^2\right] \ . \ee A
straightforward computation results in the following scalar
curvature: \be R= \frac{27}{2}\frac{S^{7/3} N(S,Q,\Lambda)}{ \left(
3\,{S}^{4/3}-{Q}^{2}-\Lambda\,{S}^{2} \right) ^{3} \left(
3\,{S}^{4/3}-5\,{Q}^{2}+\Lambda\,{S}^{2} \right) ^{2} } , \ee with
\bea N(S,Q,\Lambda)= &
&42\,{Q}^{2}{S}^{7/3}\Lambda-34\,S{Q}^{4}\Lambda-5\,{S}^{3}{Q}^{2}{
\Lambda}^{2}-18\,{Q}^{4}{S}^{1/3}\nonumber\\
&-&7\,{S}^{5}{\Lambda}^{3}+36\,{S}^{
11/3}\Lambda+15\,{S}^{13/3}{\Lambda}^{2}-162\,{S}^{3}+108\,{Q}^{2}{S}^
{5/3} \ .
\eea

From the expression for the scalar curvature it is obvious that the
singularities are located at the points satisfying the equation
$3\,{S}^{4/3}-{Q}^{2}-\Lambda\,{S}^{2} \rightarrow 0$, which
coincide with the points where $T\rightarrow 0$, and at the points
satisfying the equation
$3\,{S}^{4/3}-5\,{Q}^{2}+\Lambda\,{S}^{2}\rightarrow 0$, which are
the points where $C_Q\rightarrow \infty$. This proves that the
curvature scalar correctly reproduces the thermodynamic behavior in
the grand canonical ensemble.

It would be interesting to prove explicitly the invariance of the above results 
with respect to a total Legendre transformation, i.e., when the thermodynamic 
potential is the Gibbs potential $G(T,\phi) = M-TS - \phi Q$, which is also associated
with the grand canonical ensemble  \cite{cgk00}. Unfortunately, it is not possible to express $S$ and $Q$
in terms of $T$ and $\phi$ so that the Gibbs potential cannot be
written explicitly. However, in the limiting case of a vanishing
cosmological constant, it is possible to find explicitly the Gibbs
potential and, as shown in \cite{tqs12}, the
corresponding thermodynamic metric can be computed and
the invariance with respect to the total Legendre transformation can
be shown at the level of the scalar curvature.

Let us now consider the canonical ensemble with fundamental equation
(\ref{feqlm1}). The enthalpy $H$ is in this case the thermodynamic
potential and the coordinates of the equilibrium manifold are
$E^a=(S,\phi)$. The corresponding thermodynamic metric is obtained
from
 (\ref{gdown}) as
\be g=\left(1-\frac{3}{4}\phi^2-\frac{\Lambda}{3}S^{2/3}\right)
\left[\frac{4}{27}S^{-2/3}\left(1-\frac{3}{4}\phi^2+\frac{\Lambda}
{3}S^{2/3}\right)dS^2 - S^{4/3}d\phi^2\right] \ , \ee which leads to
the curvature scalar \be R = \frac{N(S,\phi,\Lambda)}{(12S^{1/3} -
9\phi^2S^{1/3} - 4\Lambda S)^3 (12S^{1/3} - 9\phi^2S^{1/3} +
4\Lambda S)^2}\ , \ee with \be N(S,\phi,\Lambda) = (12S^{1/3} -
9\phi^2S^{1/3} - 4\Lambda S)^2 + \Lambda
S^{4/3}[27\phi^2(4-3\phi^2)+2\Lambda S^{2/3}(9\phi^2-4\Lambda
S^{2/3}-4)]\ . \ee The curvature singularities which follow from the
limit
$12S^{1/3} - 9\phi^2S^{1/3} + 4\Lambda S\rightarrow 0$
determine the phase transition structure of the black hole, because
they coincide with the divergences of the specific heat $C_\phi$.
The second set of singularities for which  $12S^{1/3} -
9\phi^2S^{1/3} - 4\Lambda S\rightarrow 0$ corresponds to the limit
$T\rightarrow 0$ and indicates the break down of the thermodynamic
description of the black hole. This shows that the thermodynamic
behavior of this black hole which follows from the canonical
ensemble is correctly reproduced in GTD.

\section{Conclusions}
\label{sec:con}

In this work, we analyzed the problem of ensemble dependence in the
context of GTD. In the Euclidean path-integral approach it is known
that the stability properties of a black hole can depend on the
statistical ensemble. In the thermodynamic limit, a change of
ensemble can be simply performed as a Legendre transformation that
acts on the thermodynamic potential. It then follows that the
thermodynamic properties of a black can depend on the choice of
thermodynamic potential.

On the other hand, the invariance with respect to Legendre
transformations plays an essential role in GTD. So far, the metrics
used in GTD can be either invariant under total Legendre
transformations ($G^I$ and $G^{II})$ or invariant also with respect
to partial transformations $(G^{III})$. The important question is
whether the ensemble dependence affects the formalism of GTD. We
have shown in this work that in fact GTD can handle correctly this
dependence and, moreover, it explains partially why in GTD a
particular thermodynamic metric must be chosen in order to correctly
reproduce the thermodynamic behavior of black holes. In fact, by
analyzing explicitly a particular black hole in the EMGB gravity
theory with cosmological constant, we showed that the phase
transition structure is not invariant with respect to partial
Legendre transformations. This implies that the metric $G^{III}$
cannot be used to describe black hole thermodynamics.

However, the family of metrics invariant under total transformations
contains two metrics, $G^I$ and $G^{II}$, but only $G^{II}$ can be
used for black holes with second order phase transitions.
Practically, the difference between $G^I$ and $G^{II}$ is only the
signature: the first one is Euclidean and the second one is
pseudo-Euclidean. The question is how such a particular distinctness
can differentiate between systems with first order and second order
phase transitions. Preliminary results seem to indicate that this
can be used to formulate an invariant definition of phase
transitions \cite{quev13}.

\section{Acknowledgments}

We would like to thank the members of the GTD-group at the UNAM for
fruitful comments and discussions. This work was supported by
CONACyT-Mexico, Grant No. 166391,  DGAPA-UNAM and by CNPq-Brazil.

{999}

\end{document}